\documentclass[preprint,showpacs]{revtex4}

\usepackage[american]{babel}
\usepackage{epsfig}
\usepackage{hyperref}

\begin{document}
\citeindextrue
\title{Improved structural ordering in sexithiophene thick films grown on single
crystal oxide substrates}
\author{C. Aruta\footnote
{Correspondence should be addressed to: carmela.aruta@na.infn.it},
P. D'Angelo, M. Barra, G. Ausanio and A. Cassinese}

\address{CNR-INFM Coherentia and Dipartimento di Scienze Fisiche,
Universit\'{a} di Napoli "Federico II", Piazzale Tecchio 80,
Napoli 80125, Italy}

\date{\today}

\begin{abstract}
We report on sexithiophene films, about 150nm thick, grown by
thermal evaporation on single crystal oxides and, as comparison,
on $Si/SiO_{2}$. By heating the entire deposition chamber at
$100^{\circ}C$ we obtain standing-up oriented molecules all over
the bulk thickness. Surface morphology shows step-like islands,
each step being only one monolayer height. The constant and
uniform warming of the molecules obtained by heating the entire
deposition chamber allows a stable diffusion-limited growth
process. Therefore, the regular growth kinetic is preserved when
increasing the thickness of the film. Electrical measurements on
differently structured films evidence the impact of the inter
island separation region size on the main charge transport
parameters.
\end{abstract}
 \maketitle

\section{Introduction}
 The increasing attention towards organic electronics has been requiring
 a deeper knowledge of the materials
and interfaces affecting electrical transport properties in
organic films \cite{FacchettiScience}. Understanding the
relationship between film morphology and charge transport is a key
issue to improve the devices performances. In particular, the
substrate surface and the growth conditions have shown a strong
influence on the structural and electronic properties of organic
films\cite{6Fichou,YuJAP2007}. Indeed, the film morphology at the
buried interfaces can be different from that in the bulk
\cite{Kline}, thus strongly affecting the charge transport in thin
film transistors. Under this respect, anisotropic linear
$\pi$-conjugated molecules like sexithiophene (T6) can be obtained
as polycrystalline or highly oriented thin films \cite{Servet,
5Ivanco}, with charge transport depending on the long range
molecular ordering \cite{7Garnier}. Because of electronic
interactions between the molecules and active surfaces,
$\pi$-conjugated molecules tend to grow in lying-mode on
relatively active surfaces such as metals \cite{Lukas, Yoshikawa,
1Kiguchi}. On the contrary, on relatively inert surfaces, such as
$SiO_{2}$ or $KBr$ substrates \cite{Hamano, Ikeda}, these
molecules can grow in the standing-mode. Lattice
incommensurability or roughness of the substrates also influence
the growth mode of $\pi$-conjugated molecules. Standing mode or
flat-lying-mode growth is observed on poly-crystalline
\cite{Okajima} or single-crystalline metal surfaces
\cite{Yoshikawa, Kiguchi}, respectively. In addition, nucleation
and growth processes strongly influence the structure and
morphology of high-vacuum evaporated long-chain
molecules\cite{Muccini}. In thick films (thickness higher than
100nm) the three-dimensional (3D) growth maintains a memory of a
layer-by-layer growth, even after the first nucleation process is
completed\cite{Biscarini_PRB}. Therefore, both layer-plus-island
(Stranski-Krastanov) and island (Vollmer-Weber) growth modes can
be observed.  The size of the 3D-grains was shown to increase upon
using increased substrate temperatures, indicating that the growth
mode is a diffusion-limited process\cite{Biscarini_PRB,
Biscarini_PRL}. Well-defined structural properties are crucial
both for fundamental studies and for fabrication of optimal
devices, but the control of the growth of thick films can be
further improved. In this work we report on highly structured
thick T6 films (about 150nm thick) grown by heating the entire
deposition chamber. Like that, both standing and lying molecules
are equally warmed, in spite of their different
surface-area/volume ratio and the growth temperature have the same
effect on the growth kinetic of both types of molecular
orientation. Hundreds nanometers thick films are particularly
important in view of possible applications, such as organic
spin-valves or Field Effect Transistors (FET) with a vertical
structure, where different substrates are also employed. We have
grown T6 films on the single crystal substrates, $Al_{2}O_{3}$
(sapphire) r-plane and MgO (100). Both substrates have similar
physical properties, such as for example the dielectric constant
(about 10) and the thermal conductivity (about 30 $W/mK$). Quite
the reverse, the in-plane lattice structures are different:
$Al_{2}O_{3}$ r-plane has the in-plane hexagonal lattice with both
$a$ and $b$ axes equal to $4.75\AA$, while $MgO$ (100) has the
in-plane square lattice with axis 4.21 $\AA$. Nevertheless, we
find similar structural, morphological and electrical properties
in T6 films grown on both substrates. Although, there has been a
considerable effort towards the control of T6 growth when in form
of very thin films, the study of thicker films on oxide substrate
can be further refined\cite{Blumstengel}. The results on single
crystal oxides are compared with films grown during the same
deposition process on $Si/SiO_{2}$(100), the typical substrate
used for electrical properties and growth processes investigation.\\
\section{Experimental}
T6 films were vacuum evaporated by a Knudsen cell with a base
pressure of 1-3x$10^{-7}$ mbar. The deposition rate and the
thickeness were monitored by a quartz oscillator, with averaged
growth rate ranging between 0.2 and 0.4\AA/sec. The film thickness
was also controlled by a TENCOR profilometer. The sample holder
allowed the simultaneous growth of two $10\times10mm^{2}$ films.
For the optimization of T6 films, we used a peculiar sample
heating procedure, consisting in keeping warm the entire chamber
at the same temperature of $100^{\circ}C$ for 24 hours, before to
proceed with the deposition. Such expedient allows a warm
environment which favors the heating of both sides of the samples.
All films were structurally and morphologically characterized.
Because of the heavy sulphur atoms in the T6 structure, it is
possible to get important information on the structural properties
by X-ray diffraction (XRD) with a conventional X-ray source. We
used the Bragg-Brentano geometry in symmetrical reflection mode,
the Cu $K\alpha$ wavelength and a graphite monochromator. In
addition, an Atomic Force Microscopy (AFM) (Digital Instruments
Nanoscope IIIa), equipped with a sharpened silicon tip with an
apical curvature radius $\leq 5 nm$, was used in tapping mode with
a rate of 1Hz, to obtain images of the surface profile under
ambient conditions. After minimizing the tip size effect by the
deconvolution on each AFM image, the three-dimensional view of the
deposits was reconstructed. The electrical properties have been
also investigated by Current-Voltage (I-V) measurements. All
measurements were performed on planar samples, in vacuum
($10^{-3}$ mbar),  by means of a Keithley 487 picoammeter. A
standard two probe technique was employed by a cryogenic probe
station. In plane silver contacts have been grown on the surface
film (top contacts), obtaining conducting channels with length
$L=100 \mu m$ and width $w=5 mm$. Finally, FET devices have been
fabricated by depositing T6 films on $SiO_{2}$ 200nm thick,
thermally grown on heavily n-doped Si gate electrode, with
interdigitated gold source and drain contacts (bottom contact).
These electrodes provide a channel length $L=40 \mu m$ and a
channel width $w=20 mm$.

\section{Results and discussion}
A typical XRD spectrum of the optimized T6 film grown on
${Al_{2}O_{3}}$ substrate is reported in fig.1(a). We can index
the reflections by using the low-temperature single crystal data.
Such a structure corresponds to the $P2_{1}/n$ space group with
$a=44.708\AA$, $b=7.851\AA$ and $c=6.029\AA$\cite{6Horowitz}. Only
the ($h00$) reflections of the T6 compound can be observed.
Therefore, T6 films on single crystal oxides result very well
aligned with the long axis along the direction perpendicular to
the substrate surface. The noticeable result is the presence of
high order reflections up to (34,0,0) and the very narrow rocking
curve of the high-order reflection (20,0,0) reported in fig.1(b),
with a Full Width at Half Maximum (FWHM) about $0.2^{\circ}$.
Typical values of the rocking curve, which are considered
extremely low for an organic film, are of the order of
$1^{\circ}$\cite{Oehzelt}, though the measurements were performed
by synchrotron radiation on lower order reflections. Our
structural data demonstrate the improved crystal quality of T6
films grown in a warmed environment, being the structural ordering
propagated along the whole 150nm thickness. The corresponding AFM
images are reported in fig.2. Domains of about one micron large
can be observed in fig.2(a), which are composed by step-like
circular aggregates. The cross-section reported in fig.2(c) of the
smaller AFM scan size of fig.2(b) reveals that the step height is
about one molecule. Together with the XRD results, we can infer
that each terrace is very well structured, with molecules
vertically arranged. XRD data of T6 films grown on $Si/SiO_{2}$
substrate reported in fig. 3(a) show that the molecules
prevalently orient with the long axis perpendicular to the
substrate surface also on $Si/SiO_{2}$, in agreement with the
results reported in literature\cite{5Ivanco}. However, in such a
case we can detect minor order reflections in the $\Theta-2\Theta$
spectra [fig. 3(a)], up to the (22,0,0) only. The FWHM of the
rocking curve of the (20,0,0) T6 peak reported in fig. 3 (b) is
about $0.5^{\circ}$. The presence of minor order reflections and
the larger rocking curve in the case of $Si/SiO_{2}$ with respect
to the $Al_{2}O_{3}$ substrate, reveal the minor crystal quality
of T6 films on $Si/SiO_{2}$. It has been reported that molecules
perpendicular and parallel to the substrate can be simultaneously
present in  T6 films with submonolayer coverage grown on silicon
dioxide\cite{Loi}. Such an interface disorder will prevent the
growth of a well ordered T6 film, in agreement with our findings.
XRD results are confirmed by the AFM measurements reported in
fig.4, which reveal smaller size domains with a non homogeneous
distribution in case of T6 films on $Si/SiO_{2}$. Regular
step-like islands are occasionally observed [fig. 4(b)]. On the
contrary, we find high crystal quality and equivalent morphology
in both $Al_{2}O_{3}$ and $MgO$ single crystal oxides. We can
tentatively ascribe the good crystal quality of our thick films
grown at $100^{\circ}C$ to the uniform growth temperature of the
whole film thickness. Indeed, by heating all the chamber, the
surface of the film is always at the same temperature during the
growth process. On the contrary, in case of conventional heating
of the film from the back of the substrate, the heat transfer
changes by increasing the film thickness and the surface of the
growing film could not be at the ideal temperature during all the
deposition process\cite{Cassinese}. It has been reported that
islands made of standing molecules can progressively incorporate
flat-lying molecules during the evolution of the growth process
\cite{Dinelli}. The appearance of standing molecules in thick
films is suggested to be a result of a gradual deterioration of
the ordering of the lying molecules occurred during the increasing
of the film thickness\cite{Ivanco_SurfSc}. On the contrary, our
structural and morphological data show that the bulk of the film
is formed by ordered vertical standing molecules. However, Kowarik
\textit{et al.} \cite{Kowarik} found that, as a general feature of
the growth of organics, the molecules prefer the lying-down
orientation when the van der Waals interaction on the stepped
sapphire surface is important. In our case, the uniform heating by
the warmed chamber increases the mobility of the molecules,
favoring the standing-up orientation. Theoretical calculations of
Kubono \textit{et al.}\cite{Kubono} demonstrates that standing
molecules increase faster than lying molecules when the substrate
temperature is higher. Therefore, we can guess that the uniform
heating further increases this effect. Indeed, the growth
temperature have the same effect on the growth kinetic of both
standing and lying molecules which are equally warmed, in spite of
their different surface-area/volume ratio. As a matter of fact,
the lattice match is not the principal factor governing the
organic mechanism of growth. A further evidence of this aspect is
given by the employment of different single crystal substrates,
such as $MgO$, which give also rise to well oriented molecules.
The differences between the single crystal oxides and the
$Si/SiO_{2}$ substrates can be related to the different surface
chemical composition\cite{Dinelli} and/or surface crystal
order\cite{Ivanco_SurfSc}, thus changing the surface energy. The
growth process is controlled not only by the chemical interaction
between the molecule and the substrate surface
\cite{Biscarini_PRB} but also by the kinetics mediated by the
substrate surface order, with the latter dominating in weakly
interacting systems\cite{Ivanco_SurfSc}. The silicon oxide layer
on our $Si/SiO_{2}$ substrates is not well structured, as
confirmed by the absence of any silicon oxide reflection by XRD
and a smoother substrate surface by AFM. On the contrary,
$Al_{2}O_{3}$ substrates have edges along the crystal directions
which act as nucleation centers. It is well established that,
because of the strong electrostatic field
at step edges, isolated molecules absorb preferentially there.\\
In order to complete our study, the electrical properties of the
T6 films have been investigated both by bulk I-V measurements and
FET techniques. In any case, the measurements have been performed
in darkness and in vacuum (typically $10^{-3}$ mbar) to strongly
reduce the effects of ambient doping (i.e. $O_{2}$ and $H_{2}O$).
Bulk I-V measurements have been carried out only in the case of T6
films deposited on single crystal oxide substrates, since
undesired leakage currents were observed at high fields on
$Si/SiO_{2}$ substrates. For all investigated samples, at low
voltages ($V <100 Volt$, $E< 10^{4} V/cm$) I-V curves show a
linear behavior according to the basic Ohm's law. Conductivity
values extracted in this current range showed to be strongly
affected by the film morphology, as demonstrated by the comparison
between I-V curves for films grown in the warmed chamber and at
$100^{\circ}C$ (fig.5). Indeed, the XRD spectra of the two films
(inset of fig.5) reveal an appreciable improvement in the case of
films grown in the warmed chamber. Furthermore, in T6 films grown
at room temperature, few grains differently oriented are also
present, as confirmed by the detection of other reflections
different from the $(h00)s$ in the XRD spectra. Anyway, I-V
measurements in fig.5 show that, despite the worst structural
organization, the conductivity values (from $5 \cdot 10^{-8} S/cm$
to $6 \cdot 10^{-7} S/cm$) of films deposited at room temperature
are higher than those extracted for films grown at $100^{\circ}C$
(between $5 \cdot 10^{-9} S/cm$ and $3 \cdot 10^{-8} S/cm$). At
higher voltages, all bulk I-V measurements follow a super-linear
trend which can be well fitted by a square power law. By assuming
that in this range the current is limited by space charge effects
(Space Charge Limited Current regime) in presence of uniformly
distributed traps \cite{Horowitz_AdMat}, the I-V characteristics
can be modelled by a basic formula valid for in-plane transport
measurements\cite{Geurst}:

\begin{displaymath} I=\left(\frac{2
}{\pi}w\cdot\mu_{B}\cdot\varepsilon_{0}\cdot\varepsilon_{r}\cdot\frac{V^{2}}{L^{2}}
\right)\;
\end{displaymath}

where $\epsilon_{r}$ is the T6 dielectric permittivity (about 3)
and $\mu_{B}$ is the bulk mobility related to the free carries in
the whole film. By using this formula to fit the experimental
measurements, we have obtained that room temperature bulk mobility
$\mu_{B}$ of the charge carriers, down-scaled by the ratio between
free and trapped carriers \cite{Lampert}, ranges between $5 \cdot
10^{-5}$ and $1 \cdot 10^{-4} cm^{2} /Volt \cdot sec$ for T6 films
grown in warmed environment. In the same way, $\mu_{B}$ has been
estimated to be about $5 \cdot 10^{-2} cm^{2} /Volt \cdot sec$ for
films grown at room temperature. The better I-V performances of
the films grown at room temperature, seem to suggest that the
packed structure related to the small grain morphology better
supports the charge carrier bulk transport, if compared to the
step-islands morphology, where the presence of large boundary
regions can reduce the occurrence of conduction percolating
paths\cite{Yang}.\\
I-V curves measured at different temperatures for a T6 film
deposited on $Al_{2}O_{3}$ in warmed environment are reported in
Fig.6. It is worth to mention that both bulk mobility and
conductivity extracted by I-V curves obey a standard Arrhenius law
(inset of fig.6) with an activation energy $\Delta_{B}$ very close
to 0.2 eV in the range between 295 K and 335 K.\\
 FET measurements have been performed on T6 films with thickness
 ranging between 20 nm ad 150 nm grown both in warmed and
room temperature environment. In any case, the responses of our T6
transistors exhibit some typical features of the electrical
transport regimes dominated by charge carriers trapping
phenomena\cite{Stallinga,Torsi}. As shown in fig.7, where the
output characteristics, $I_{DS}$ \textit{vs} $V_{DS}$
(drain-source current versus voltage) curves at fixed $V_{G}$
(gate voltage) of a T6 FET device grown in warmed environment are
reported, the currents slightly decay in the heart of the
saturation region ($V_{G}\sim V_{DS}$)\cite{Stallinga}. Similarly,
the trapping effects provide the hysteresis in the measured
transfer curves ($I_{DS}-V_{G}$ at fixed $V_{DS} = -5Volt$) where
(see inset of fig.7) the current in the direct gate voltage scan
from +20 (off state) to -50 V (on state) is always higher than
that of the inverse scan (from -50 to +20 V, where the device
turns off). In this work, the mobility of free charge carriers
($\mu_{FET}$) has been simply evaluated by the slope of the direct
scan of the transfer curve in the linear regime ($V_{G}\gg
V_{DS}$), according to the basic formula:

\begin{displaymath} g_{m}=\frac{\partial I_{DS}}{\partial V_{G}}=\frac{w C_{i} \mu_{FET}}{L}\;
\end{displaymath}

where $C_{i}$ (about $17 nF/cm^{2}$) is the insulator capacitance
per unit area. Following this procedure, $\mu_{FET}$ values
between 1.1 and 1.5$\cdot 10^{-2} cm^{2}/Volt \cdot sec$ have been
typically estimated at room temperature for T6 films grown in warm
environment. No clear thickness dependence of mobility has been
evidenced in our study, supporting the basic vision that the field
effect involves only a few nanometers thick region at the
interface between dielectric and organic film. These results and
the extracted mobility values are in good agreement with previous
reports\cite{Dinelli2}. On the contrary, for T6 films grown in
room temperature environment, $\mu_{FET}$ always resulted lower
than 1$\cdot 10^{-3} cm^{2}/Volt \cdot sec$ and with less
reproducibility among the different samples. Finally, also for T6
measurements, we found that mobility is thermally activated, with
typical activation energies $\Delta_{FET}$ between 60 and 90 meV.
The large discrepancy between the electrical behavior and the
related physical parameters checked in bulk and FET measurements
are necessarily due to different electron transport phenomena
involved, which primarily reflect the different energy state
density affecting the charge carriers motion occurring in bulk and
interface regions\cite{Tanase}. Furthermore, as stated above, I-V
and FET measurements are performed on films grown on different
substrates and with slightly different morphologies. In order to
better address these fundamental issues, further studies, on both
field effect and vertical devices of T6 films grown on single
crystal oxides, are envisaged.

\section{conclusions}

We have shown that T6 thick films grown on single crystal oxides
by heating the whole chamber at $100^{\circ}C$, have a noticeable
good crystal quality all over the thickness of about 150nm. We can
explain such a result in terms of surface temperature of the film
which is constant during the whole process and uniform along the
whole molecule, thus favoring standing-up orientation. At our
growth conditions, the localized substrate/molecule interactions
occurring in the early growth stages determine the crystal quality
of the whole film, even in the case of such sizeable thickness.
Indeed, minor crystal quality is obtained for T6 films on
$Si/SiO_{2}$ substrate. Such a difference respect to the single
crystal oxide substrates can be associated to the surface crystal
ordering. Density of nuclei depends upon the nature of the
substrate and affects the earlier stage of growth when 3D-islands
start to form. In particular, step-edges at the single-crystal
oxide surfaces act as stable anchoring sites for the molecules.
Successively, other arriving molecules diffuse on the surface
until they anchor to the previous ones. Such a
diffusion-controlled growth depends on the mobility of molecules
and it is favored by the uniform and constant growth temperature
we used. The grain boundaries between large step-like islands with
well oriented molecules limit the long-range electrical in-plane
transport, as demonstrated by our electrical measurements.
However, well-defined anisotropic structural features are crucial
for fundamental study of the anisotropic charge transport in such
a long-chain $\pi$-conjugated molecules. The control of the growth
process and the physical properties of thick films will be useful
to better understand the role of the interface as well as of the
bulk. This can be relevant for the design of devices based on
vertical structure such as
rectifying diodes and organic spin valves.\\
\section{Acknowledgments}
We acknowledge stimulating discussions with C. Albonetti, F.
Biscarini and E. Lunedei. Fruitful discussions of the XRD data
with A. Geddo-Lehmann are also acknowledged. 
\clearpage

\begin{figure}
\includegraphics{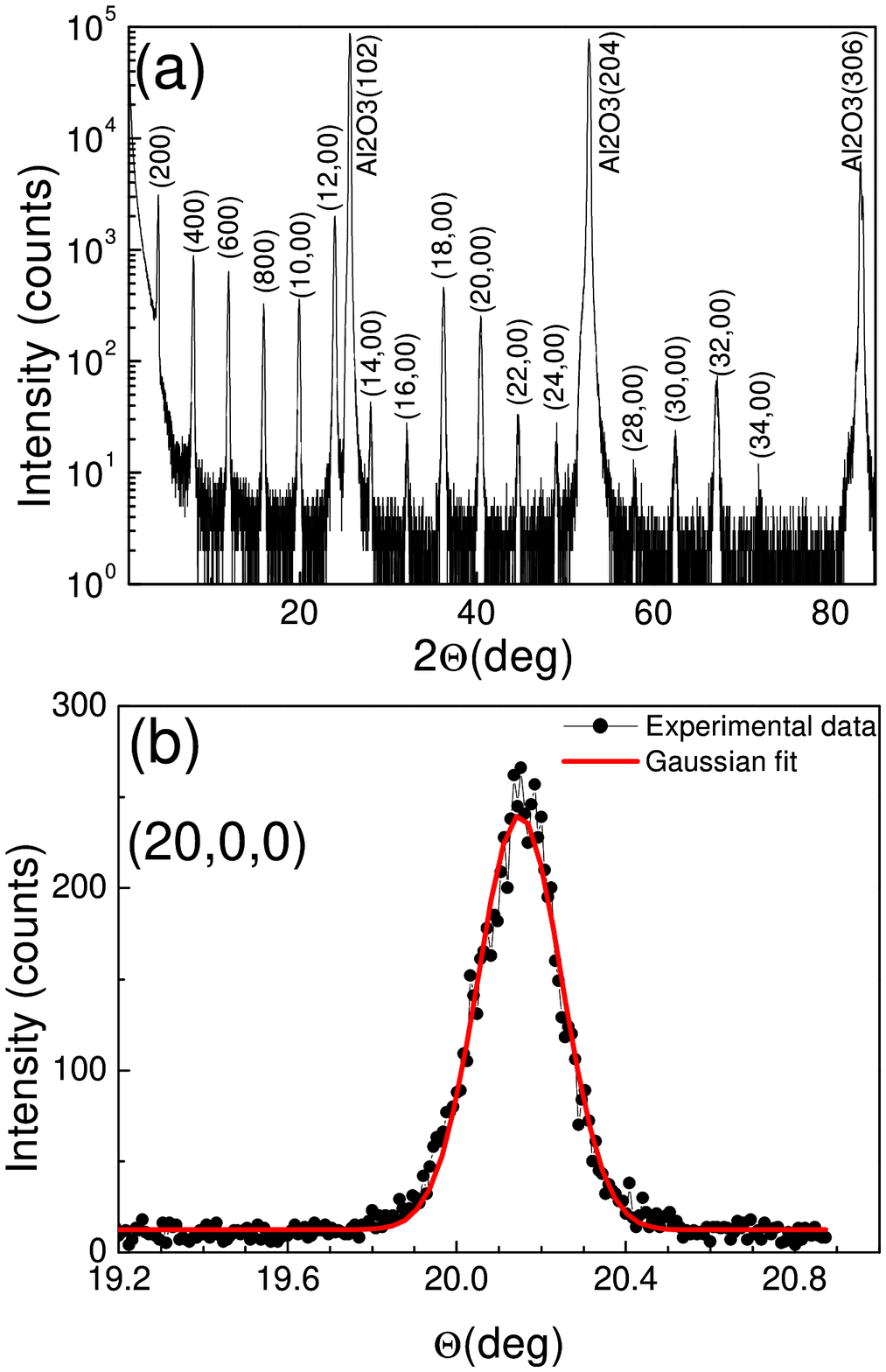}
\caption{(Color online) (a) XRD measurements of an optimized T6
film grown at $100^{\circ}C$ on $Al_{2}O_{3}$ substrate. (b)
Rocking curve of the (20,0,0) reflection.}
\end{figure}
\clearpage

\begin{figure}
\includegraphics{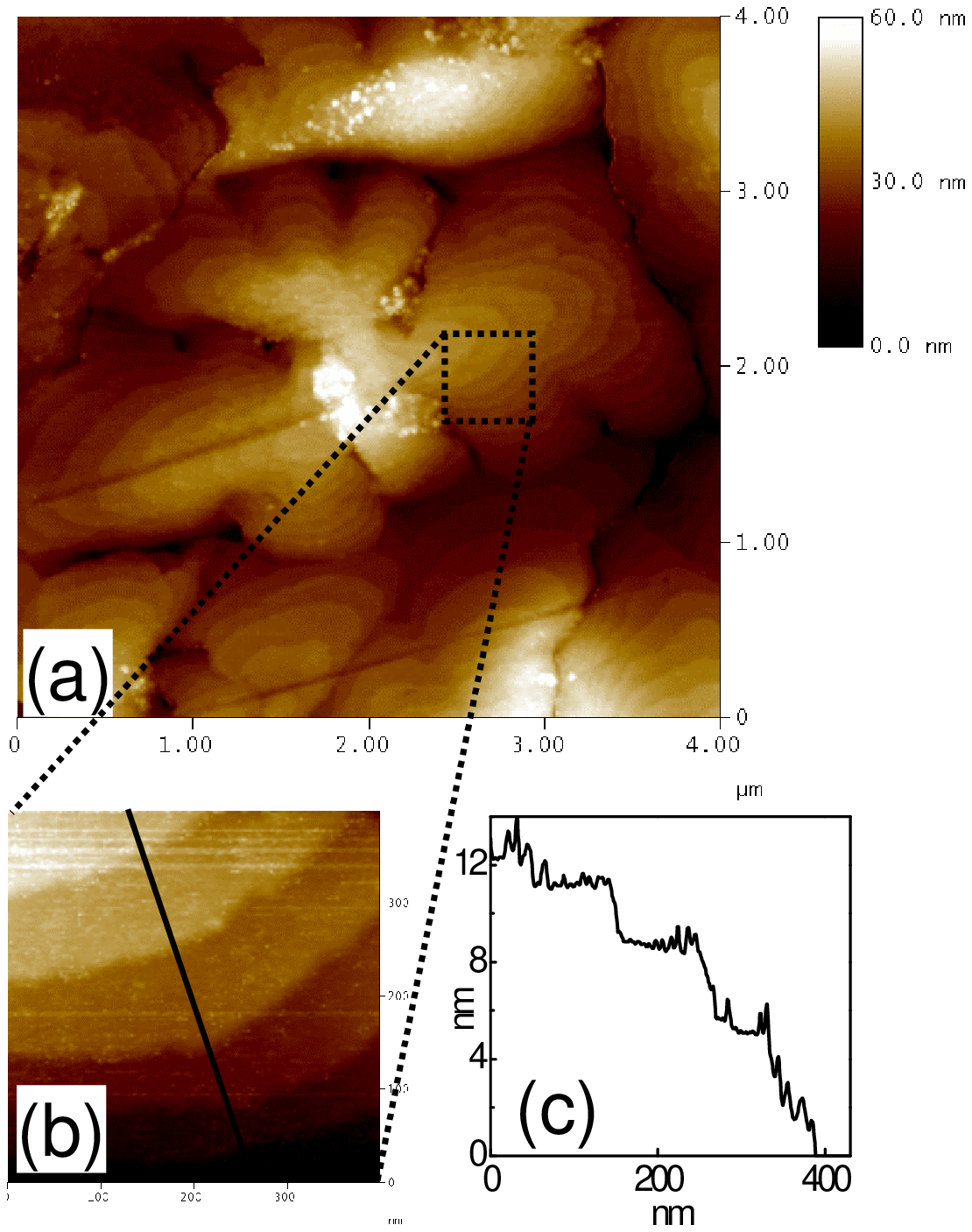}
\caption{(Color online) AFM images of an optimized T6 film grown
at $100^{\circ}C$ on $Al_{2}O_{3}$ substrate: (a)scan size of
$4\times4\mu m$ and Z-range $0-60 nm$, (b) scan size of
$430\times430 nm$ and Z-range $0-20 nm$, (c) cross section along
the straight line of panel (b).}
\end{figure}
\clearpage

\begin{figure}
\includegraphics{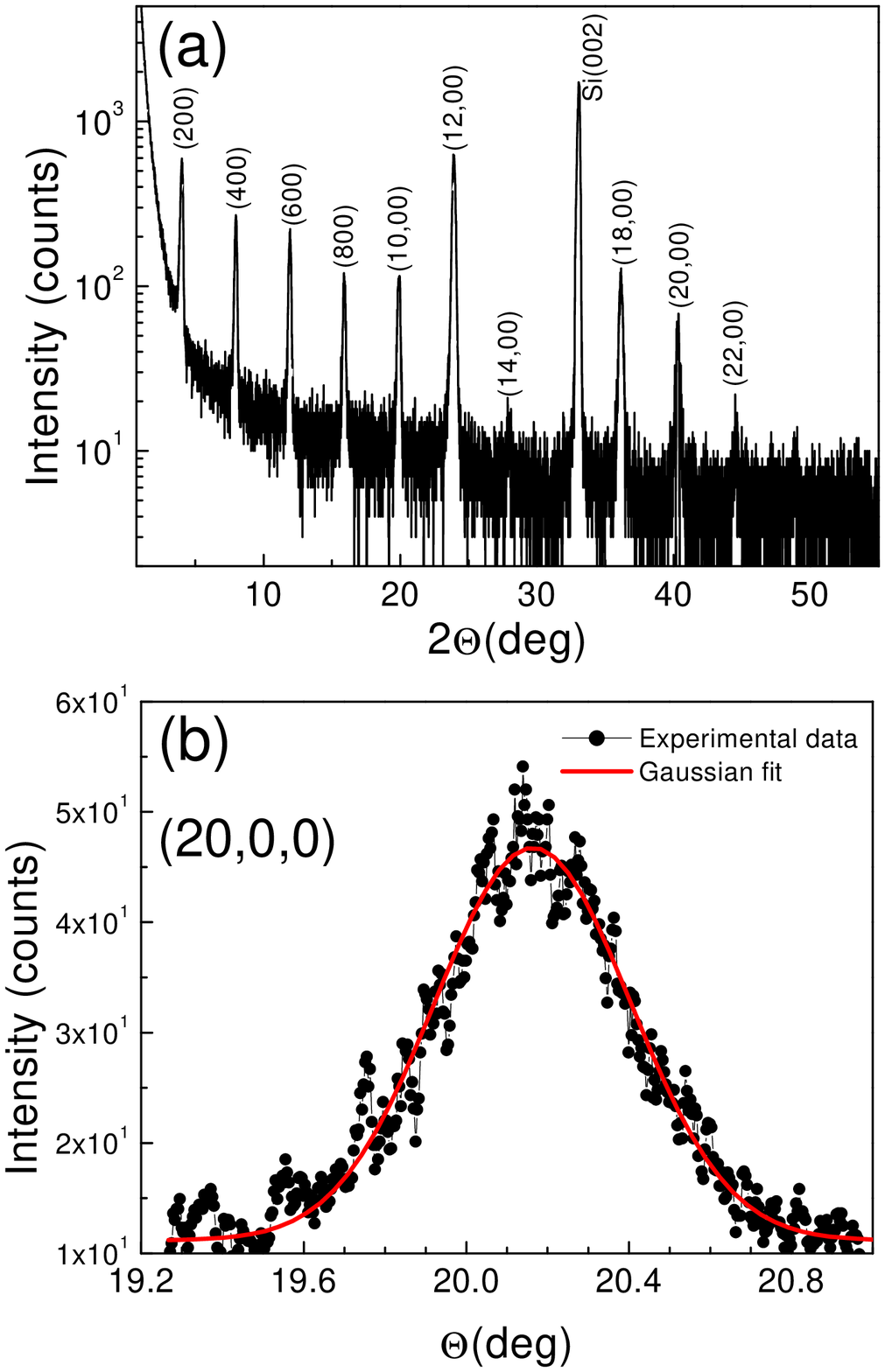}
\caption{(Color online) (a) XRD measurements of a T6 film grown at
$100^{\circ}C$ on $Si/SiO_{2}$ substrate. (b) Rocking curve of the
(20,0,0) reflection.}
\end{figure}
\clearpage

\begin{figure}
\includegraphics{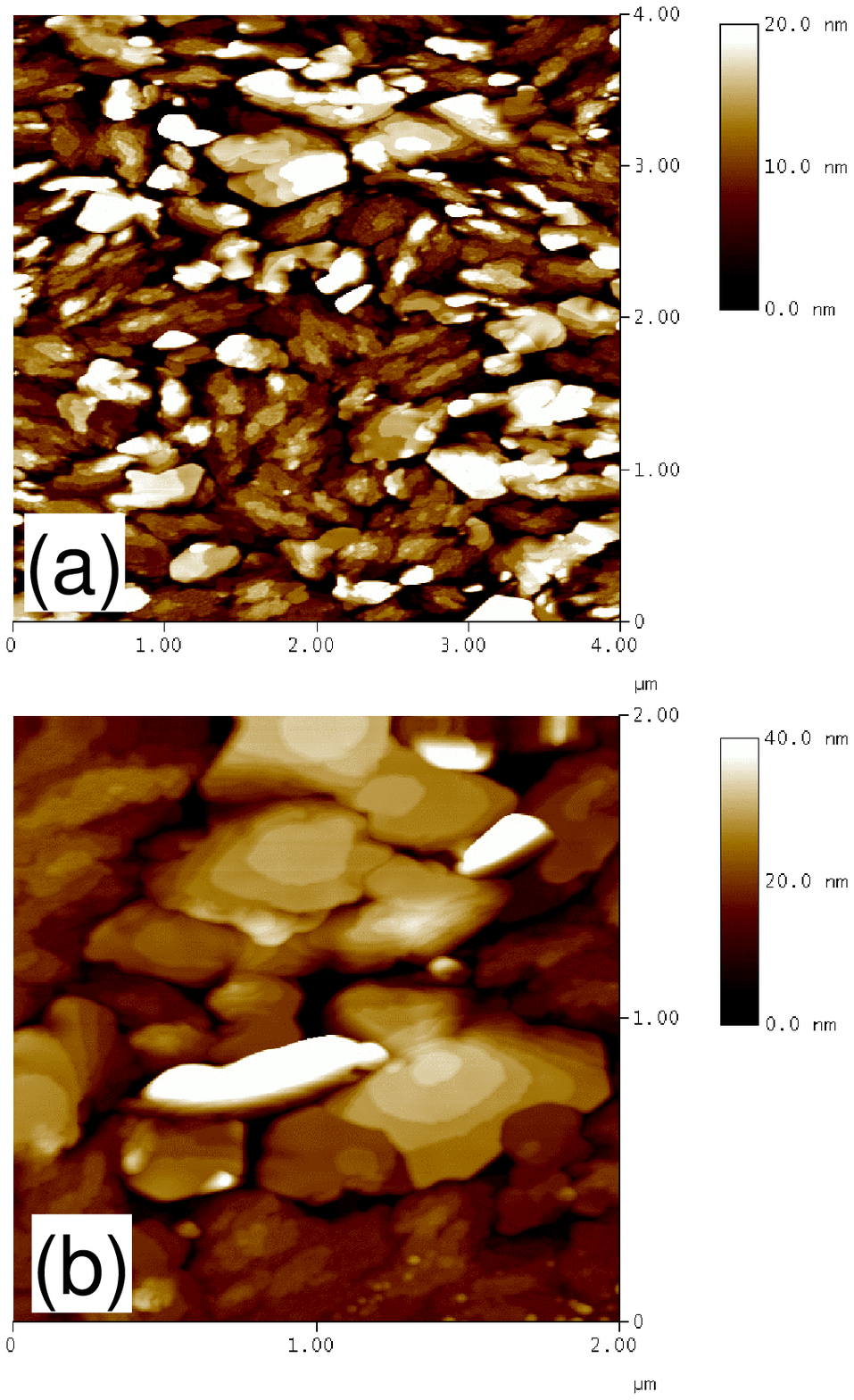}
\caption{(Color online) AFM images of a T6 film grown at
$100^{\circ}C$ on $Si/SiO_{2}$ substrate: (a) scan size of
$4\times4\mu m$ and (b) scan size of $2\times2\mu m$. }
\end{figure}
\clearpage

\begin{figure}
\includegraphics{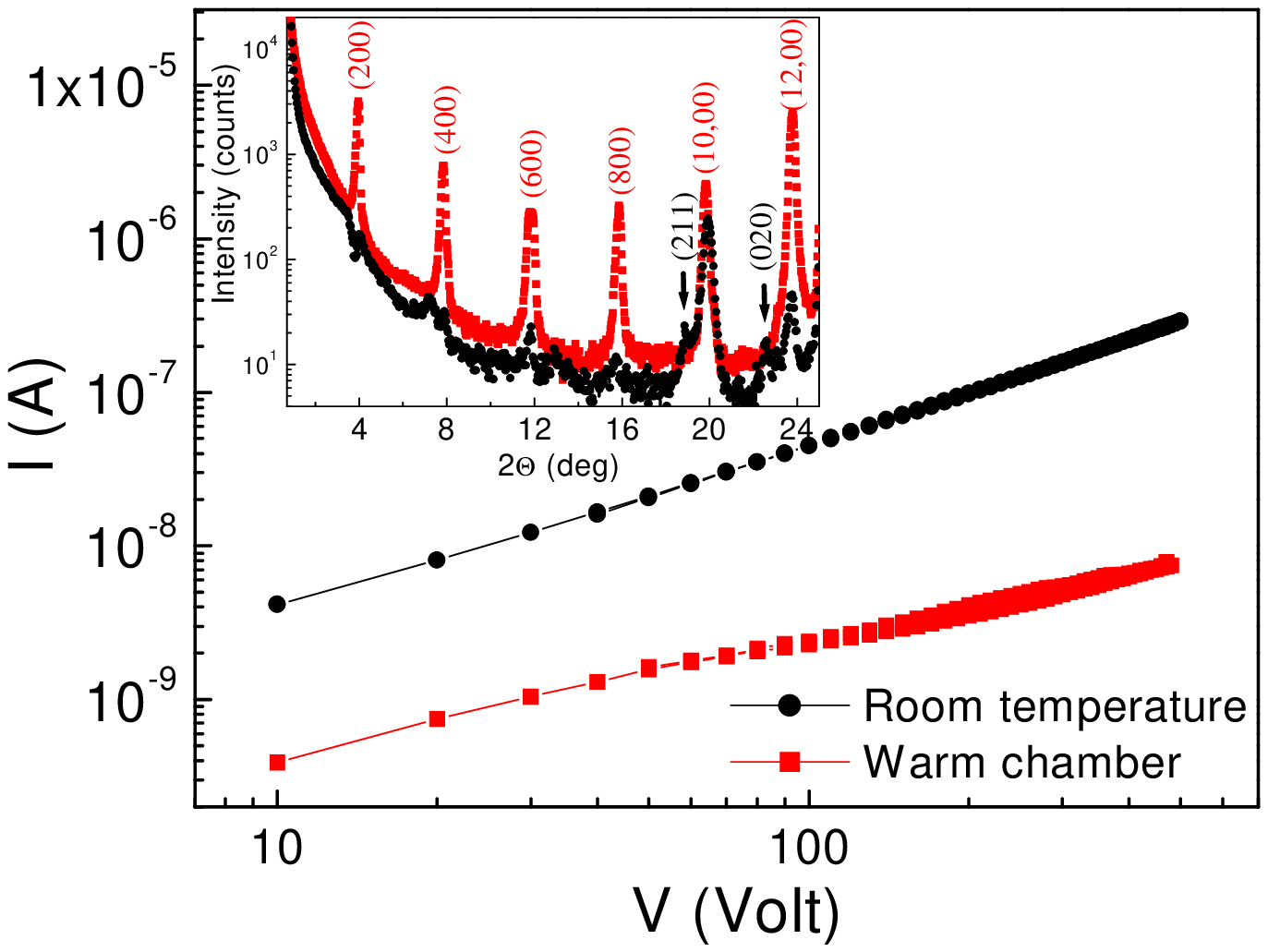}
\caption{(Color online) Bulk I-V measurements of T6 films
deposited at room temperature and in the warmed environment on
$Al_{2}O_{3}$ substrate. In the inset, the comparison of the
corresponding XRD spectra is also reported.}
\end{figure}
\clearpage

\begin{figure}
\includegraphics{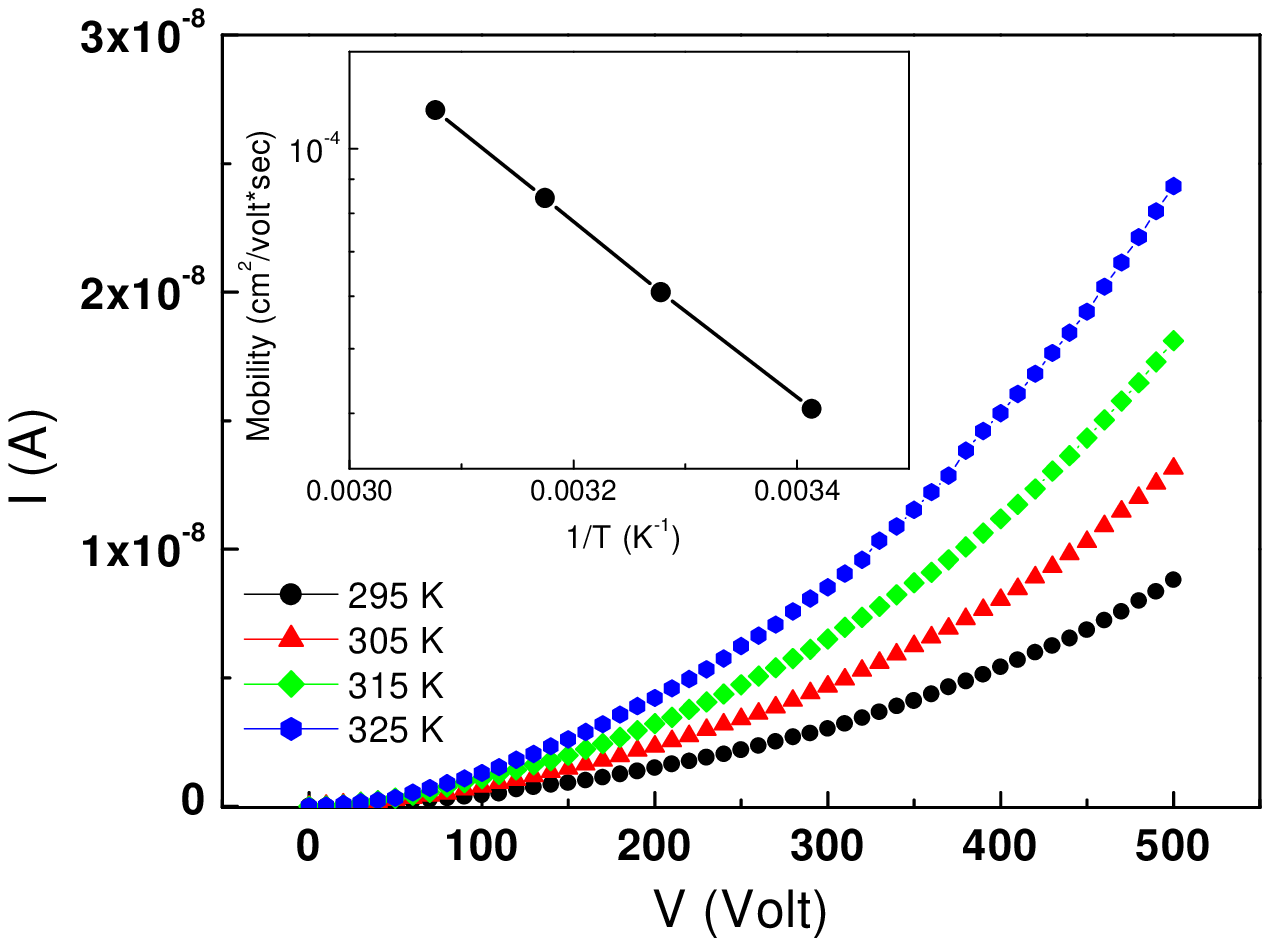}
\caption{(Color online) Bulk I-V measurements at different
temperatures for a T6 film deposited on $Al_{2}O_{3}$ substrate.
The mobility as a function of the inverse temperature is reported
in the inset.}
\end{figure}
\clearpage

\begin{figure}
\includegraphics{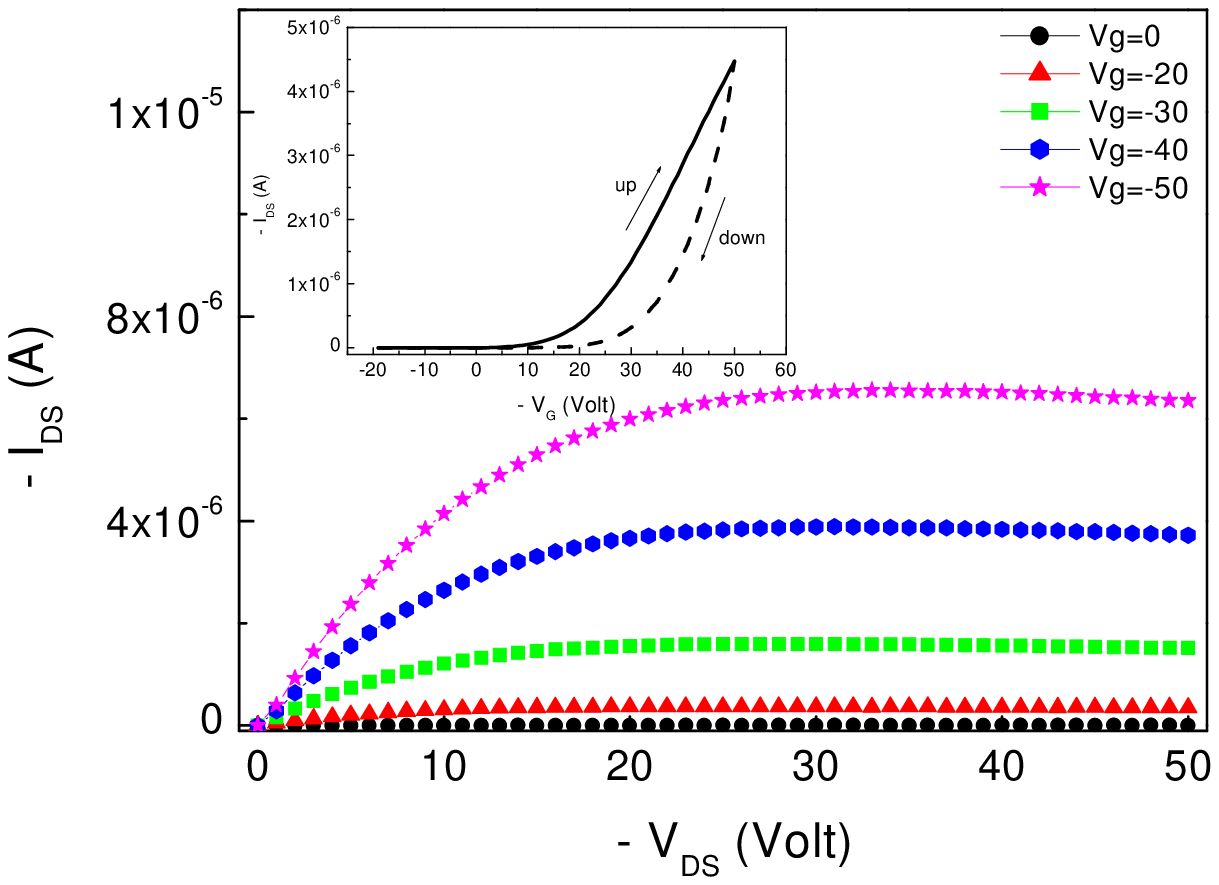}
\caption{(Color online) $I_{DS}$ - $V_{DS}$ curves for a T6 FET
device at various gate voltages ($V_{G}$). In the inset the
transfer curve measurement at $V_{DS}=-5 Volt$ is reported. }
\end{figure}

\end{document}